\documentclass[prb,twocolumn,notitlepage,10pt]{revtex4-1}
\usepackage[colorlinks=true,linkcolor=blue,citecolor=blue]{hyperref}
\usepackage{graphicx,amsmath,bm,amssymb}

\begin{document}
\title{Self-optimized construction of transition rate matrices from accelerated atomistic simulations with Bayesian uncertainty quantification}
\author{Thomas D Swinburne}
\author{Danny Perez}
\affiliation{Theoretical Division T-1, Los Alamos National Laboratory, Los Alamos, NM, 87545, USA}
\begin{abstract}
  A massively parallel method to build large transition rate matrices from temperature accelerated molecular dynamics trajectories is presented.
  Bayesian Markov model analysis is used to estimate the expected residence time in the known state space, providing crucial uncertainty quantification
  for higher scale simulation schemes such as kinetic Monte Carlo or cluster dynamics. The estimators are additionally used to optimize where
  exploration is performed and the degree of temperature acceleration on the fly, giving an autonomous, optimal procedure to explore the state
  space of complex systems. The method is tested against exactly solvable models and used to explore the dynamics of C15 interstitial defects in iron.
  Our uncertainty quantification scheme allows for accurate modeling of the evolution of these  defects over timescales of several seconds.
\end{abstract}
\maketitle

The vast size and complexity of the potential energy landscape of materials make the investigation of their long-time dynamical evolution extremely difficult, as significant free energy barriers between different regions of configuration space prohibit the use of direct simulation methods. Indeed, molecular dynamics (MD) simulations of materials are typically restricted to sub-microsecond timescales, a time that is often much too short for a trajectory to cross the barriers that determine the long-time behavior. This makes extrapolation of long-time behavior based on short simulations fraught with danger.

Overcoming the extremely restrictive timescale limitation of MD is an longstanding challenge and numerous solution strategies have been proposed over the years. In {\em open-ended} situations where to goal is to generate dynamically correct evolution from a given initial condition without regards to possible end states, these methods often adopt one of two philosophies. First, trajectory-based methods such as accelerated molecular dynamics (AMD)\cite{voter1997,voter1998parallel,TAD,perez2009} and adaptive kinetic Monte Carlo \cite{henkelman2017,beland2011} generate individual trajectories that span long timescales without having to extensively explore configuration space. They do so by breaking up the problem of generating a long trajectory into that of generating a proper sequence of state-to-state transitions, which can be effectively be carried out using specifically crafted MD simulations. The second class of techniques, including methods such as Discrete Path Sampling\cite{wales2002discrete} or Markov State Models\cite{pande2010everything}, instead begin by thoroughly exploring the energy landscape, thereby producing a kinetic model that can then be post-processed to infer long-time behaviors.

While the local nature of the exploration required by the first class of approaches typically lead to more accurate and affordable results, it produces only one (or a few) of an astronomically large number of possible trajectories; the representativity of the results it generates can therefore be difficult to assess. On the other hand, the second approach produces a comprehensive {\em global} model of the dynamics that can account for the contribution of large ensembles of trajectories, but the accuracy of its prediction requires that the underlying model be complete (or at least, "sufficiently" so), an assumption that can be hard to assess, as fully sampling configuration space is typically impossible for non-trivial systems. Quantifying the completeness of models of the potential energy landscape has therefore recently emerged as a critical issue \cite{jain2013,chill2014bench,ghiringhelli2015big}. It is important to note that this same challenge also affect trajectory-based methods that rely on having a complete {\em local} description of the landscape (e.g., as in adaptive kinetic Monte Carlo \cite{henkelman2017,beland2011}).
A further challenge that has received comparatively less attention is that generating a sufficiently complete model that is accurate enough to make long-time predictions
is likely to be an extremely computationally costly endeavor. Finding optimal strategies to allocate computational resources, in particular on massively parallel architectures, can therefore be expected to be paramount in making such approaches practical and scalable.

\begin{figure}[!t]
  \includegraphics[width=\columnwidth]{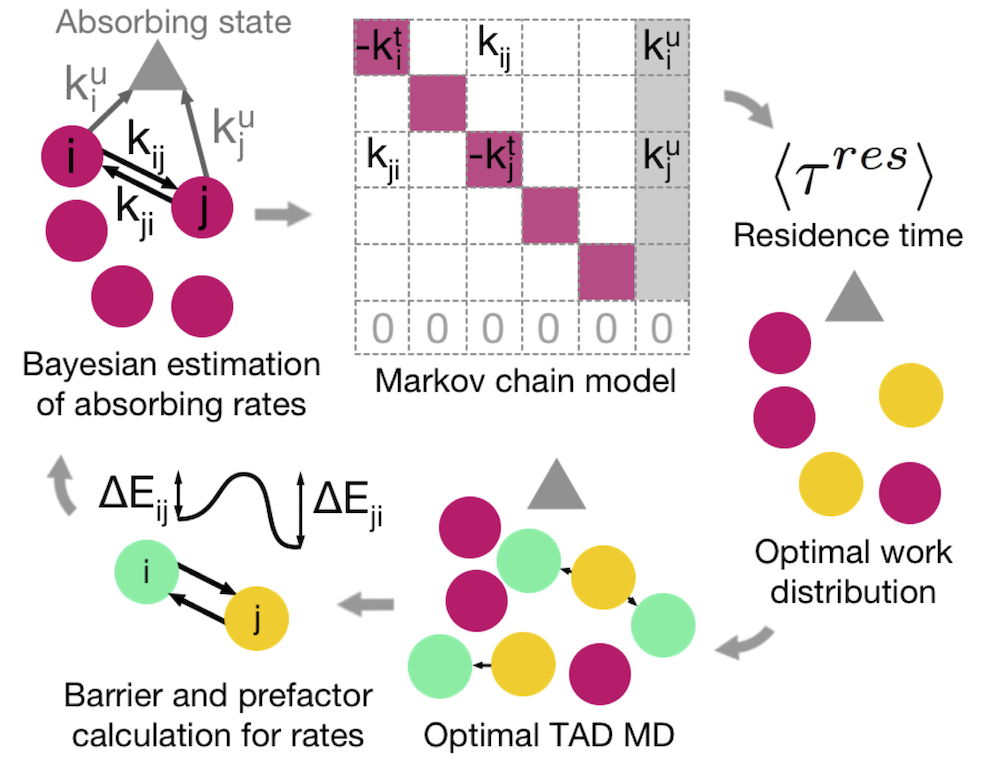}
  \caption{TAMMBER workflow. TAD MD produces interstate transition trajectories which are analyzed by Bayesian rate estimators and static calculation. An absorbing Markov chain gives then gives the expected residence time and optimally allocates resources and the degree of temperature acceleration. The cycle is then repeated until the target residence time is achieved.}
  \label{rmfig}
\end{figure}

In this paper, we introduce a self-optimizing scheme called TAMMBER (Temperature Accelerated Markov Models with Bayesian Estimation of Rates) that comprehensively address these challenges. As illustrated in Fig. \ref{rmfig}, TAMMBER relies on an AMD method, namely temperature accelerated dynamics (TAD)\cite{TAD,zamora2016},
as an efficient local exploration tool. The local completeness of the TAD  exploration is assessed using a Bayesian framework .
TAMMBER then invokes the mathematics of absorbing Continuous Time Markov Chains (CTMC)
\cite{novotny,boulougouris2005monte,boulougouris2007dynamical,chill2014,bhoutekar2017new,chatterjee2015uncertainty} to provide a global exploration completeness metric,
the expected {\it residence time} in the known configuration space. This completeness metric is then systematically optimized using a parallel
resource allocation protocol.

To put the central concepts of this paper in a concrete setting, consider a system with a total discrete state space $\mathcal{S}$. States are here defined as
basins of attraction under energy minimization, as is customary for hard materials.
After a given period of exploration with TAD, we will have discovered a subset $\mathcal{K}\subset\mathcal{S}$ of the total state space, the known states.
Whilst an observed system state $i\in\mathcal{K}$ will be connected to a subset of states $\mathcal{S}_i\subset\mathcal{S}$, in general we will
have observed only a subset of connections $\mathcal{K}_i\subset\mathcal{K}$ in the explored state space\footnote{We note that in general all possible connections
in $\mathcal{K}$ will not have been observed, i.e.  $\mathcal{K}_i
\neq\mathcal{S}_i\cap\mathcal{K}$}. Defining the transition rate from a state $i$ to a state $j$ at a temperature ${\rm T}=1/({\rm k_B\beta})$ as $k_{ij}(\beta)$, the {\it total} escape rate for a state $i$ reads
\begin{equation}
    k^{\rm tot}_i(\beta) \equiv \sum_{j\in\mathcal{S}_i} k_{ij}(\beta).
\end{equation}
As discussed above, due to incomplete exploration we will only have access to the {\it observed} escape rate
\begin{equation}
  k^{\rm obs}_i(\beta)\equiv \sum_{j\in\mathcal{K}_i} k_{ij}(\beta), \label{run_tr}
\end{equation}
which immediately defines the statewise {\it unknown} escape rate
\begin{equation}
  k^{\rm un}_i(\beta) \equiv k^{\rm tot}_i(\beta) - k^{\rm obs}_i(\beta) = \sum_{j\in\mathcal{S}_i\setminus\mathcal{K}_i} k_{ij}(\beta).
\end{equation}
where $\mathcal{S}_i\setminus\mathcal{K}_i\equiv\{x:x\in\mathcal{S}_i\,,\, x\notin\mathcal{K}_i \}$ is the set difference between $\mathcal{S}_i$ $\mathcal{K}_i$. In an absorbing CTMC, the unknown rates $k^{\rm un}_i$ are encoded as transition rates to single or multiple absorbing states (sinks) that represents the entire
unexplored space and unobserved connections within the explored state space. Standard results\cite{ethier2009} can be used to obtain the {\em residence time} of the model,
which quantifies the expected amount of time before an unknown transition should statistically occur. The residence time
can be interpreted as a typical duration over which model trajectories are a valid representation of the true system trajectories, providing
an important uncertainty quantification metric when using the calculated rate matrices in coarse grained methods such as kinetic Monte Carlo or cluster dynamics. The direct optimization of this metric
with respect to additional computational work then provides an optimal allocation strategy to maximally improve the quality of the model at the smallest possible computational cost.
Upon completion of a batch of TAD simulations, the model is updated and the cycle repeats.

The mathematics of absorbing CTMC have previously been used to accelerate kinetic Monte Carlo simulations of superbasin escape\cite{novotny} and highly heterogeneous
glassy systems\cite{boulougouris2005monte,boulougouris2007dynamical} though in both of these cases the the chains were fully specified and this partitioning
into two groups was made for computational convenience. Estimation of the unknown rate for each state has previously been investigated in molecular dynamics simulations
of biological systems\cite{bhoutekar2017new,chatterjee2015uncertainty}, whilst high temperature dynamics has also been used to estimate the degree of sampling
completeness in individual states\cite{chill2014} which is closely related to estimation of the unknown rate.
The central novelty of this work is both the robust form of our estimators for the unknown escape rate from each state and an expression for the expected decrease in
the unknown rate with additional computational work. Using these expressions we are able to determine both the optimal degree of temperature acceleration
for each state on the fly and the response of the residence time to additional computational effort applied to a given {\it distribution} of states,
an essential feature for application to massively parallel computers. 
Importantly,
by optimizing the distribution of computational resources to grow the residence time as fast as possible, we optimize a {\it global} metric of sampling completeness, a point we return to below.

The paper is organized as follows. In section \ref{secTAD} we recall the temperature accelerated dynamics method\cite{TAD} and detail how the method may be extended to allow for a variable high temperature. In section \ref{secRATES} we derive a novel Bayesian estimators for the $k_{ij}(\beta)$ of observed transitions ($j\in\mathcal{K}_i$) at any desired temperature and the {\it unknown} escape rate $k^{\rm un}_i(\beta)$ from each state. In section \ref{optTAD} we derive an analytical expression to determine the state-wise optimum temperature to reduce the unknown rate for each state and use these results to derive the residence time and optimal control
protocol using an absorbing CTMC in section \ref{secAMC}. Details of the numerical implementation are described in \ref{secTAM}, along with a test against known rate
matrices (using kinetic Monte Carlo to generate trajectories) and a demonstrative study of C15 interstitial defects in iron.

\section{Temperature Accelerated Dynamics}\label{secTAD}
The temperature accelerated dynamics (TAD) method \cite{TAD,zamora2016} is an AMD technique that exploits the Poisson distribution of rare event escape times\cite{Gesu2017} and the
approximations of harmonic transitions state theory (HTST)\cite{Kramers} to generate statistically-correct low temperature trajectories from high temperature MD data alone. When the transition barriers
are sufficiently large, TAD can provide a very significant acceleration of the state-to-state dynamics as compared to MD, because the first event to occur at low temperature will typically occur after only a
much shorter time at a higher temperature. TAD provides a statistically sound way of assessing when the said first event has indeed been observed at high temperature, and hence of selecting a proper
low-temperature transition.

We recall that when the free energy barrier $\Delta F_{ij}$ for some state transition $i\to{}j$ is much larger than the thermal energy $\beta^{-1}$, the transition rate $k_{ij}(\beta)$
is well approximated by the Arrhenius expression\cite{Kramers}
\begin{equation}
  k_{ij}(\beta) = \omega_{ij}\exp[-\beta\Delta F(\beta)] \simeq \nu_{ij}\exp[-\beta\Delta E_{ij}]. \label{HTST}
\end{equation}
The second equality in Eq.\ (\ref{HTST}) constitutes the HTST approximation,
where the entropic contribution to the barrier $\Delta S_{ij}$ is assumed to be constant, leading to a constant prefactor $\nu_{ij}=\omega_{ij}\exp(\Delta S_{ij}/{\rm k_B})$ and a potential energy barrier $\Delta E_{ij}$. The extension of the approach developed here to incorporated anharmonic entropic effects\cite{swinburne2017} will be the topic of a future publication.
HTST  (\ref{HTST}) can be exploited in the present context by noting that the event times for a Poisson process of rate $k(\beta)$ are distributed as
\begin{equation}
  \tau_{ij}(\beta) \sim -\log|\mathcal{U}(0,1)|/k_{ij}(\beta),
\end{equation}
where $\mathcal{U}(0,1)$ is the uniform distribution on the unit interval; from this functional form it is clear that a valid event time $\tau_{ij}(\beta')$ at a different temperature can be obtained from a sample $\tau(\beta)$ through
\begin{equation}
  \tau_{ij}(\beta') = \tau_{ij}(\beta)\frac{k_{ij}(\beta)}{k_{ij}(\beta')} \simeq \tau_{ij}(\beta)\exp\left[(\beta'-\beta)\Delta E_{ij}\right], \label{TADT}
\end{equation}
where the HTST approximation was used to obtain the final relation. As $\Delta E_{ij}$ is readily calculated using minimum energy path algorithms such as the NEB method\cite{neb},
after a process has been observed for the first time at high temperature, we can thus generate a corresponding first passage times at other temperatures.

%
In TAD, this remapping of first passage times is exploited as follows.
Consider a state $i$ that has dynamically accessible pathways to a set of connected states $j\in\mathcal{S}_i$,
with escape rates $k_{ij}(\beta) = \nu_{ij} \exp[-\beta\Delta E_{ij}]$. TAD  uses high temperature MD to produce high temperature escape times $\{\tau_{ij}(\beta_{\rm H})\}$
to a subset of connected states $\mathcal{K}_i\subset\mathcal{S}_i$. Once an escape is detected, the system is put back into state $i$, accumulating a total effective state time $\tau_i(\beta)$. The escape times along each pathways can then be rescaled to yield a set of low temperature first passage times $\{\tau_{ij}(\beta_{\rm L})\}$, which will in general have a different ordering given the nonlinear character of (\ref{TADT}). In conventional TAD, the goal is to identify the transition that should have occurred first, i.e., the transition which corresponds to the minimum value of $\tau_{ij}(\beta_{\rm L})$.
The central difficulty is the observed escape times are only to a subset of all possible final states $\mathcal{K}_i$.
It is therefore important to avoid prematurely choosing a low-temperature transition from the set transitions so far observed at high temperature.
TAD achieves this through a Poisson uncertainty bound; defining a minimum prefactor $\nu_{\rm min} \simeq 0.1$THz, high temperature MD is carried out until the probability that the
proper first escape pathway at low temperature has yet to be observed at high temperature is less than $\delta\sim0.05$.
The worst possible case in this setting is that of a low barrier and low prefactor process with rate $\nu_{\rm min}\exp(-\beta_{\rm H}E^{\rm min}_i)$, where $E^{\rm min}$ is the smallest barrier
that could potentially remain unobserved after running dynamics at high temperature for a time $\tau_i(\beta_{\rm H})$.
It is simple to show that\cite{zamora2016}
\begin{equation}
  E^{\rm min}_i = \beta^{-1}_{\rm H}\log\left[\frac{\nu_{\rm min}\tau_i(\beta_{\rm H})}{\log(1/\delta)}\right],\label{emin}
\end{equation}
which produces a low temperature effective state time
\begin{equation}
  \tau_i(\beta_{\rm L}) = \tau_i(\beta_{\rm H})\exp[(\beta_{\rm L}-\beta_{\rm H})E^{\rm min}_i],\label{state_time}
\end{equation}
after which we have a confidence $1-\delta$ to have seen all
relevant first passages up to this time.

\begin{figure}
  \includegraphics[width=\columnwidth]{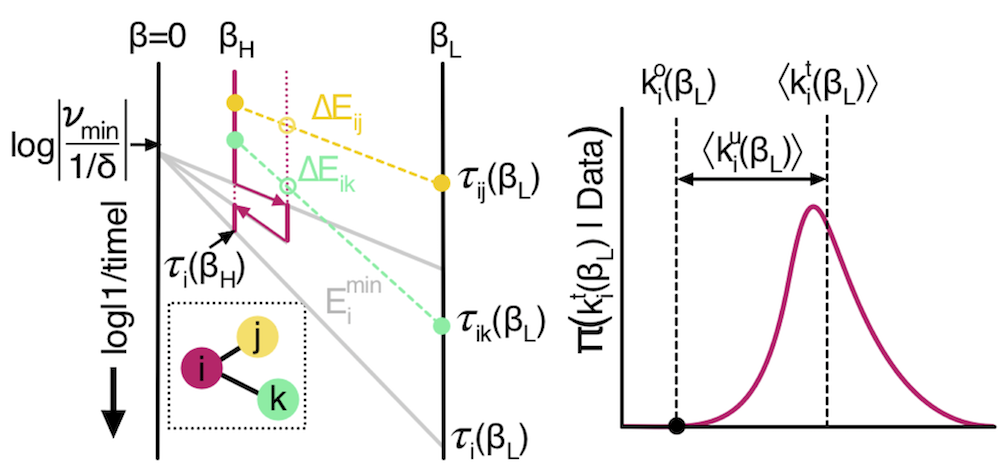}
  \caption{Left: Illustration of the TAD method developed here. Low temperature first passage times become valid as they are swept past, whilst the high temperature can be changed to accomodate trajectory data at a new temperature. Left: Qualitatively representative posterior for the total escape rate from a state. The unknown rate is the difference between the mean total rate and the observed rate.}
  \label{tadfig}
\end{figure}

In the original TAD method, the goal is to follow the first valid escape process, i.e. state time is accumulated until $\tau_i(\beta_{\rm L})$ is greater than the smallest rescaled first passage time.
In the present case we continue accumulating state time, producing an ever greater catalogue of valid low temperature escape times (i.e., all of those whose rescaled event times are smaller than $\tau_i(\beta_{\rm L})$),
for use in our rate estimators detailed in the next section. As the total state time $\tau_i$ and first passage times $\tau_{ij}$ are defined at any temperature,
we can incorporate multi-temperature data by using (\ref{state_time}).
An illustration of this procedure is detailed in Fig.\ \ref{tadfig}.
\section{Determination of the known and unknown escape rates from a state}\label{secRATES}
In order to apply the absorbing CTMC analysis which is central to our approach, we need to produce an estimate for the individual rates $k_{ij}(\beta)$
between known states at any given temperature and for the {\it unknown} escape rate $k^{\rm un}_i(\beta)$ from each known state.
In the following we derive Bayesian likelihood estimators for the individual and total escape rates from a given state using the first passage trajectories $\tau_{ij}(\beta)$ and state time $\tau_i(\beta)$.

\subsection{Estimation of individual escape rates}
Once an individual escape process from a state $i$ to a state $j$ has been observed, the NEB method can be used to obtain the minimum energy pathway and
hence the energy barriers $\Delta E_{ij}$ and $\Delta E_{ji}$. To calculate the individual escape rates $k_{ij}$ and $k_{ji}$ we therefore only require calculation of the rate prefactors $\nu_{ij}$ and $\nu_{ji}$.

It is possible to directly calculate an estimate for the rate prefactors using harmonic transition state theory\cite{Kramers}. A key advantage is that the HTST approximation to $\nu$ is
often accurate and produces a rate matrix which satisfies detailed balance, but calculation requires computationally expensive diagonalization of the Hessian matrix
at the end points and saddle point of the transition pathway\cite{huang2013}. 
An alternative approach is to directly estimate the rate prefactor from the transitions observed during MD simulation.
A disadvantage of this approach is that this requires multiple observed transitions to give reliable results and that the resultant prefactors have no guarantee of satisfying detailed balance.
Nevertheless, when transitions are sufficiently rapid (which can be expected when using accelerated approaches such as TAD) sufficient data can often be obtained to
produce accurate estimates.

In this section we derive a simple Bayesian estimator for the rate prefactor which incorporates prior knowledge of the prefactor and dynamical information from an ensemble of escape-replace trajectory data. The prior estimate for the prefactor can either be set to a typical value of $\nu_0=1$THz or a static HTST calculation. In a Bayesian setting, this knowledge can be encoded in an unnormalized prior distribution
\begin{equation}
  \pi_0(\nu_{ij}) = \exp[-\alpha(\nu_{ij}/\nu_0-1)^2/2],\label{prior}
\end{equation}
where $\nu_0$ is the prior estimate and $\alpha$ will turn out to control the number of data points that are needed to override the influence of the prior. As a result, if a full HTST calculation is undertaken, $\alpha$ should be large as we are confident that our prior is accurate. In practice, as a full prefactor calculation is computationally intensive, we only undertake such calculations when we expect dynamical data to be rare, i.e. when $\Delta E_{ij}$ is large, though many strategies can be envisaged, for example performing an approximate calculation with the degree of approximation reflected in the prior distribution.

We represent escape-replace trajectory data as $\{\beta_i,\tau_i,N_{ij}\}$, where $\beta_i$ is the inverse temperature, $\tau_i$ is the total effective state time at that temperature and $N_{ij}$ is the total number of $i\to{}j$ transitions observed\footnote{The generalization to multi-temperature data is straightforward and allows an additional estimate $\Delta E_{ij}$}. For clarity of presentation we also define the dimensionless, Boltzmann scaled trajectory times
\begin{equation}
  \tilde{\tau}_{i;j} = \tau_i\nu_0\exp[-\beta_i\Delta E_{ij}],
\end{equation}
where the notation distinguishes $\tilde{\tau}_{i;j}$ from the first passage times $\tau_{ij}$. Using the Poisson likelihood for $N$ events in a time $\tau$, $(k\tau)^N\exp(-k\tau)/N!$, the HTST relation (\ref{HTST}) and the prior distribution (\ref{prior}), the unnormalized posterior for the rate prefactor reads
\begin{equation}
  \pi(\nu_{ij} | \tilde{\tau}_{i;j},N_{ij})
  =  \pi_0(\nu)(\nu_{ij}\tilde{\tau}_{i;j})^{N_{ij}}
  \exp(-\nu_{ij}\tilde{\tau}_{i;j}/\nu_0).
\label{PosteriorIndividual}
\end{equation}
Whilst the posterior distribution is quite cumbersome, we can produce an estimator for $\nu_{ij}$ using the maximum log likelihood (MLL) technique, where the logarithm of the unnormalized posterior (\ref{PosteriorIndividual}) is maximized with respect to $\nu_{ij}$, a well known procedure in parameter estimation\cite{myung2003tutorial}. Through elementary operations one obtains from $\partial_\nu\log\pi = 0$ a quadratic equation for $\nu_{ij}$ which has the unique positive solution
\begin{equation}
  \nu_{ij} = \frac{\nu_0}{2}\left[
  1-\frac{\tilde{\tau}_{i;j}}{\alpha}
  +
  \sqrt{\left(
  1-\frac{\tilde{\tau}_{i;j}}{\alpha}
  \right)^2+4\frac{N_{ij}}{\alpha}}
  \right]. \label{mll_pf}
\end{equation}
In the small time and data limit $\tilde{\tau}_{i;j}\ll \alpha, N_{ij}\ll\alpha$, we find $\nu_{ij}=\nu_0$, as one would expect, whilst at long times $\tilde{\tau}_{i;j}\gg \alpha$ we recover $\nu_{ij} = N_{ij}\exp[\beta_i\Delta E_{ij}]/\tau_i$, which the minimum variance estimator for this Poisson process\footnote{We note that this minimum variance estimator will systematically overestimate the rate, as $P\left(\frac{N}{\tau}\exp(\beta\Delta E) > \nu\right) = {\Gamma(N,N)}/{\Gamma(N)} > 1/2$}.

We have found $\alpha\simeq10$ to give robust sampling behavior using a standard initial prefactor $\nu_0=0.1$THz. A key advantage of the Bayesian approach is that if a more detailed HTST prefactor calculation is undertaken to give a more reliable prior estimate, we make the prior distribution sharper by increasing the $\alpha$ parameter. As a result, a much larger amount of dynamical data is required to significantly change the posterior prediction of the prefactor, thus naturally incorporating the two estimation methods.

\subsection{Estimation of the unknown escape rate from a state}

{

With calculated prefactors and energy barriers $\{\nu_{ij},\Delta E_{ij}\}$ for each observed escape process, we can readily calculate the corresponding escape rates $k_{ij}(\beta)$.
Furthermore, using the procedure described above, we can also obtain an effective state time $\tau_i(\beta)$ at any given temperature. In this section, we show how this information,
taken together with the sampled first passage time $\tau_{ij}(\beta)$ obtained with TAD, can be
used to produce a Bayesian estimator for the {\it unknown} escape rate from the generated first passage time trajectory, again at any temperature.


In anticipation of the results below, we time order the individual escapes labels such that $\tau_{i(j-1)}(\beta)<\tau_{ij}<\tau_{i(j+1)}$ and then define the running total rate
\begin{equation}
  k^{\rm obs}_{i;j}(\beta) \equiv \sum_{\tau_{ik}(\beta)\leq\tau_{ij}(\beta)} k_{ik}(\beta) \label{run_tr_tau},
\end{equation}
i.e., the running total rate $k^{\rm obs}_{i;j}(\beta)$ includes all events that occurred at times $\tau_{ij}(\beta)$ that are lower or equal to the effective residence time at $\beta$, $\tau_i(\beta)$.
As all rates are evaluated at a constant temperature for the entirety of this section, we now omit $\beta$ for clarity of presentation.

To build a posterior distribution for the unknown rate, consider the likelihood of observing a first passage $i\to{}j$ after waiting a time $\tau_{ij}-\tau_{i(j-1)}$ since the last event. For a postulated total rate $k$, the remaining rate in this time interval is simply $k-k^{\rm obs}_{i;j-1}$, giving a likelihood for $\tau_{ij}$ of
\begin{align}
  \pi(\tau_{ij}|k^{\rm tot}_i = k) &= [k-k^{\rm obs}_{i;j-1}](\tau_{ij}-\tau_{i(j-1)})  \label{lk}\\
  &\times\exp\left(-[k-k^{\rm obs}_{i;j-1}](\tau_{ij}-\tau_{i(j-1)})\right).\nonumber
\end{align}
We note the use of the remaining total rate in the interval $[\tau_{i(j-1)},\tau_{ij}]$ is essential to give the correct likelihood. In addition, as we know that the total rate satisfies $k^{\rm tot}_i = k^{\rm obs}_i + k^{\rm un}_i$, we can write the same likelihood for $\tau_{ij}$ for a postulated unknown rate $k$ as

\begin{align}
  \pi(\tau_{ij}|k^{\rm un}_i = k) &= [k+k^{\rm obs}_{i}-k^{\rm obs}_{i;j-1}](\tau_{ij}-\tau_{i(j-1)})\label{lku}\\
    &\times\exp\left(-[k+k^{\rm obs}_{i}-k^{\rm obs}_{i;j-1}](\tau_{ij}-\tau_{i(j-1)})\right),\nonumber
\end{align}
where $k^{\rm obs}_{i}$ (as defined in (\ref{run_tr})) is the sum of the escape rates for events at any temperature, independent of their first passage times at $\beta$.
A total likelihood for all the observed event times for a postulated unknown rate is simply the product of (\ref{lku}) for each event satisfying $\tau_{ij}(\beta)\leq\tau_i(\beta)$, multiplied by the likelihood of not seeing any other events over a time $\delta\tau^W_i=\tau_i - \max_{\tau_{ij} < \tau_i}(\tau_{ij})$ to give
\begin{equation}
  \pi(\{\tau_{ij}\}|\tau_i,k^{\rm un}_i = k) = \exp(-k\delta\tau^W_i)\prod_{\tau_{ij}\leq\tau_i}\pi(\tau_{ij}|k^{\rm un}_i = k)\label{tlku}
\end{equation}

We can now use Bayes' formula to construct an unnormalized posterior for the unknown rate, using the Jefferies prior\cite{jeffreys} $\pi_0(k) = 1/k$ for the initial likelihood function $k\exp(-kt)$. Removing all multiplicative factors independent of the postulated unknown rate, as these will disappear under renormalization, we obtain a central result of this paper, an unnormalized posterior distribution for the unknown rate
\begin{align}
    \pi(k^{\rm un}_i | \tau_i, \{\tau_{ij}\})
     &=
     \frac{\exp\left(-k^{\rm un}_i\tau_i\right)}{k^{\rm un}_i+k^{\rm obs}_i}\nonumber\\
     &\times\prod_{\tau_{ij}<\tau_i}\left(k^{\rm un}_i+k^{\rm obs}_i-k^{\rm obs}_{i;j-1}\right)
     .\label{posterior_tk}
\end{align}

We emphasize that although all trajectory information is used in the individual rate calculations, we only use first passage information in the Bayesian posterior (\ref{posterior_tk}).
It can in fact be shown that subsequent passages in fact do not contribute additional information, as we assume that the rate for a given process can be calculated once it has been observed.
This is ideal for implementation in a TAD setting, as multi-temperature MD data can be incorporated to produce an effective first passage trajectory at a wide range of desired temperatures.

A prediction for the unknown rate $\langle k^{\rm un}_i(\beta) \rangle$ and total rate $\langle k^{\rm un}_i(\beta)\rangle$ at an inverse temperature $\beta$, can now be produced by evaluating (\ref{posterior_tk}), yeilding moments
\begin{equation}
  \langle [k^{\rm un}_i(\beta)]^n\rangle =
  \frac{\int_0^\infty k^n\pi(k |\beta,\tau_i,\{\tau_{ij}\}){\rm d}k}{\int_0^\infty \pi(k |\beta, \tau_i,\{\tau_{ij}\}){\rm d}k}, \label{tr_mom}
\end{equation}
where we have reintroduced the temperature dependence explicitly. In appendix \ref{app:anal} we show that these integrals can be expressed analytically by exploiting properties of exponential integrals
 and a recursive scheme to expand the product, avoiding numerical quadrature issues.

This is the first important result of this manuscript: the first moment, namely the mean, will be used as an estimator of the unknown rate out of a given state given an observed sequence
of first passage times generated with TAD. This provides a crucial local completeness metric.
The higher moments also prove critical to solve the important question of the choice of the optimal high temperature at which the TAD procedure should be carried out
in order to maximize computational efficiency, a problem which we discuss next.


\section{Optimal TAD temperature}\label{optTAD}
The TAD method uses an elevated temperature ${\rm T_H} = 1/({\rm k_B}\beta_{\rm H})$ to reduce the computational effort required to produce a valid set of first passage times
and pathways at some lower temperature ${\rm T_L} = 1/({\rm k_B}\beta_{\rm L})$. When all barriers are sufficiently large compared to ${\rm k_B}{\rm T_L}$, the efficacy of TAD
method initially increase with increasing ${\rm T_H}$ away from ${\rm T_L}$. However, if ${\rm T_H}$ becomes too high, transitions with very large energy barriers will
become more frequent. As characterizing these transitions incurs a cost but contribute very little to the low temperature total rate, the computational efficiency of the procedure should ultimately decrease with increasing ${\rm T_H}$.
In addition, known events will reoccur more frequently at higher ${\rm T_H}$, increasing the frequency which the system must be re-prepared in the initial state in order to accumulate additional effective state time.

These arguments indicate that there will in general exist an optimum high temperature ${\rm T_H}$, the precise value of which depends on the desired outcome. Recent work\cite{shim2011adaptive} has investigated finding the optimal ${\rm T_H}$ in TAD to produce a single valid escape event from a given state, i.e. a single rescaled first passage time less than the effective state time ($\tau_{ij}(\beta_{\rm L})<\tau_i(\beta_{\rm L})$).
In this section, we instead ask for the temperature which maximizes the decrease of the expected low temperature unknown rate $\langle k^{\rm un}_i(\beta_{\rm L})\rangle$ with respect to additional computational
effort $c_i(\beta_{\rm H})$ that consists in carrying out the TAD procedure at temperature $\beta_{\rm H}$, namely
\begin{equation}
  \beta^{\rm TAD}_i = \arg\max_{\beta_{\rm H}}\left[-\frac{{\rm d}\langle k^{\rm un}_i(\beta_{\rm L})\rangle}{{\rm d} c_i(\beta_{\rm H})}\right]
\end{equation}

Given that the simulation cost is dominated by force calculation (a.k.a., force calls), the total computational effort per unit high temperature MD time can be written in units of force calls as
\begin{equation}
  \frac{{\rm d} c_i(\beta_{\rm H})}{{\rm d} \tau_i(\beta_{\rm H})} = \dot{c}_{\rm MD} + c_{\rm ST} k^{\rm obs}_i(\beta_{\rm H}) + c_{\rm NEB}k^{\rm un}_i(\beta_{\rm H}),
\end{equation}
where $\dot{c}_{\rm MD}$ is the number of force calls per unit MD time in frequency units, $c_{\rm ST}$ is the cost of state identification and preparation in force calls and $c_{\rm NEB}$ is the cost of a NEB calculation in force calls. In a typical example, where transition rates are quoted in THz and the MD timestep is a femtosecond, we have $\dot{c}_{\rm MD}=1000$, $c_{\rm ST}\simeq1000$ and $c_{\rm NEB}\simeq10000$. By the chain rule we make the useful expansion
\begin{equation}
  \frac{{\rm d}\langle k^{\rm un}_i(\beta_{\rm L})\rangle}{{\rm d} c_i(\beta_{\rm H})}
  =
  \frac{{\rm d}\langle k^{\rm un}_i(\beta_{\rm L})\rangle}{{\rm d} \tau_i(\beta_{\rm H})}
  \left(
  \frac{{\rm d} c_i(\beta_{\rm H})}{{\rm d} \tau_i(\beta_{\rm H})}\label{CR_C}
  \right)^{-1}.
\end{equation}

To evaluate the first term in (\ref{CR_C}) we first consider the expected change in the low temperature unknown rate from a small interval $\delta\tau_i(\beta_{\rm H})$ of high temperature MD when $\mathcal{E}$, a new transition is observed, or $!\mathcal{E}$, when no new transition occurs. The corresponding change in the low temperature state time, $\delta\tau_i(\beta_{\rm H})$, is readily evaluated through use of (\ref{emin}) as
\begin{equation}
  \delta\tau_i(\beta_{\rm L})
  =
  \delta\tau_i(\beta_{\rm H})
  \frac{\beta_{\rm L}}{\beta_{\rm H}}\left(\frac{\log(1/\delta)}{\nu_{\rm min}\tau_i(\beta_{\rm H})}\right)^{\beta_{\rm L}/\beta_{\rm H}-1}.
\end{equation}

We evaluate changes in $k^{\rm un}_i(\beta_{\rm L})$ through perturbation theory applied to expectation values over the low temperature posterior for the total rate, $\pi(k^{\rm tot}_i |\beta_{\rm L},\tau_i)$. If no event is seen in high temperature MD, the new posterior is given by
\begin{align}
  \pi(k^{\rm un}_i |\beta_{\rm L},\tau_i+\delta\tau_i)=&\pi(k^{\rm un}_i |\beta_{\rm L},\tau_i)\exp\left(-[k^{\rm un}_i]\delta\tau_i\right).
\end{align}
To leading order in $\delta\tau_i(\beta_{\rm L})$ the expected change in the unknown rate takes the simple form
\begin{equation}
  \langle\delta k^{\rm un}_i(\beta_{\rm L}) | !\mathcal{E}\rangle
  =
  -\delta\tau_i(\beta_{\rm L})\left[
  \langle [k^{\rm un}_i(\beta_{\rm L})]^2\rangle
  - \langle k^{\rm un}_i(\beta_{\rm L})\rangle^2
  \right].
\end{equation}

If an event $\mathcal{E}$ is seen in high temperature MD, to a state $p$ with a rescaled low temperature rate $k_{new} = k_{ip}(\beta_{\rm L})$, the new posterior distribution is given by
\begin{align}
  \pi(k^{\rm un}_i |\beta_{\rm L},\tau_i+\delta\tau_i)=&\pi(k^{\rm un}_i+k_{new} |\beta_{\rm L},\tau_i)
  \nonumber\\
  &\times
  \left(k^{\rm un}_i + k^{\rm obs}_i - \max k^{\rm obs}_{i;j}\right)
\end{align}
Whilst we can progress without any assumptions, to simplify the expectation value over this new distribution we take the mild assumption that $\max k^{\rm obs}_{i;j}\simeq k^{\rm obs}_i$, i.e. that the majority of the rate has been seen at the temperature of interest. We have found this to hold in practice, and can be expected from the form of the rescaled state time $\tau_i$. Under this approximation, the expected change in the unknown rate reads
\begin{equation}
  \langle\delta k^{\rm un}_i(\beta_{\rm L}) |\mathcal{E}\rangle
  =
  -k_{new}+\frac{\langle [k^{\rm un}_i(\beta_{\rm L})]^2\rangle
  - \langle k^{\rm un}_i(\beta_{\rm L})\rangle^2}{\langle k^{\rm un}_i(\beta_{\rm L})\rangle}.
\end{equation}

To complete this expression we require an estimate for the new low temperature rate $k_{new}=k_{ip}(\beta_{\rm L})$, ideally without making any additional assumptions on the spectrum of escape rates. We base our assumption on the expected first passage time relation $\langle\tau_{ip}\rangle=1/k_{new}$. New events are therefore expected to be first observed in order of descending rate.
If the barrier spectrum is dense, then a reasonable estimate for the next new event rate is simply the minimum of all the observed rates so far, $\min\{k_{ij}(\beta_{\rm L})\}$. However, if the spectrum has a large spectral gap, we would expect long periods without any new events, meaning the minimum of the seen rates could significantly overestimate the next event rate. In this long waiting time limit, it can be shown that the Bayesian estimator gives a max log likelihood unknown rate of $\langle k^{\rm un}_i(\beta_{\rm L})\rangle\sim1/\tau_i(\beta_{\rm L})$. As the new rate is expected to occur at a time $\tau_i(\beta_{\rm L})$, we see that the unknown rate estimate is expected to be a slight overestimate, i.e., our estimates tend to be conservative. Combining these two cases, our estimate for the next observed rate is therefore
\begin{equation}
  \langle k_{new}\rangle \simeq \min[\langle k^{\rm un}_i(\beta_{\rm L})\rangle \cup \{k_{ij}(\beta_{\rm L})\}].
\end{equation}
Given that the expected probability of seeing a new event in high temperature MD is simply $P(\mathcal{E}) = \delta\tau_i(\beta_{\rm H})k^{\rm un}_i(\beta_{\rm H})$ in the limit of small $\delta\tau_i(\beta_{\rm H})$, with $P(!\mathcal{E}) = 1-P(\mathcal{E})$, we can write the expected change in the low temperature unknown rate as
\begin{equation}
  \langle\delta k^{\rm un}_i(\beta_{\rm L})\rangle = P(\mathcal{E})\langle\delta k^{\rm un}_i(\beta_{\rm L}) |\mathcal{E}\rangle + P(!\mathcal{E})\langle\delta k^{\rm un}_i(\beta_{\rm L}) |!\mathcal{E}\rangle.
\end{equation}

Combining the above manipulations we can write the final objective function as
\begin{align}
  &-\frac{{\rm d}\langle k^{\rm un}_i(\beta_{\rm L})\rangle}{{\rm d} c_i(\beta_{\rm H})} =
  \left(
    \frac{{\rm d} c_i(\beta_{\rm H})}{{\rm d} \tau_i(\beta_{\rm H})}
  \right)^{-1}
  \Big[
    \langle k_{new}\rangle\langle k^{\rm un}_i(\beta_{\rm H})\rangle
    \label{obj_fun}
    \\
    &+
    \left(
      \frac{\tau_i(\beta_{\rm L})}{\tau_i(\beta_{\rm H})}
      -
      \frac{\langle k^{\rm un}_i(\beta_{\rm H})\rangle}{\langle k^{\rm un}_i(\beta_{\rm L})\rangle}
    \right)
    \left(
      \langle [k^{\rm un}_i(\beta_{\rm L})]^2\rangle
      -
      \langle k^{\rm un}_i(\beta_{\rm L})\rangle^2
    \right)
  \Big]\nonumber 
\end{align}

Whilst this expression appears complex, all relevant quantities can be readily calculated using our Bayesian estimator and the results derived above. In our numerical implementation, we find the maximum of (\ref{obj_fun}) to determine a different optimal $\beta_H$ for every state in the system. This determination is periodically refined to insure optimal performance.

\section{Absorbing Markov Chain Analysis}\label{secAMC}

In the preceding sections, we have described a scheme to estimate transition rates $k_{ij}(\beta)$ between known states $i,j\in\mathcal{K}$ and the unknown rate for each state $k^{\rm un}_i(\beta)$. We have also derived the expected change (\ref{obj_fun}) in the low temperature unknown rate $k^{\rm un}_i(\beta_{\rm L})$ with additional computational work at a temperature $\beta_H$ in order to determine the optimum temperature at which to carry out the TAD procedure.
In this section we use the estimated rates to build an absorbing Markov chain\cite{ethier2009}, giving both the expected residence time $\langle\tau^{res}\rangle$ spent in the known state space and the expected change in $\langle\tau^{res}\rangle$ as a result of additional computational effort.
As discussed in the introduction, the expected residence time $\langle\tau^{res}\rangle$ is an important {\it global} measure of sampling completeness, providing an estimate of the length of trajectories that can safely be generated from the CTMC; trajectories longer than $\langle\tau^{res}\rangle$ on the complete CTMC would have a significant probability of containing transitions that are not part of the estimated CTMC.
One should therefore avoid using the CTMC to make predictions on times that exceed $\langle\tau^{res}\rangle$.

We emphasize that $\langle\tau^{res}\rangle$ is a {\it global} metric that accounts for the wider energy landscape. This is quite distinct from a state-wise approach to uncertainty; for example, if a particular state has a high unknown rate, a state-wise approach would always demand more computational work in this state to reduce the uncertainty. However, in our global approach, work would only be done in this state if it is sufficiently frequently visited to have a significant influence on the global trajectory distribution. 


In our setting, $\langle\tau^{res}\rangle$  can be estimated as follows. Consider an
absorbing CTMC in a discrete state space $\mathcal{K}\cup\triangle$, namely the set of observed states and an absorbing state $\triangle$, as illustrated in figure \ref{rmfig}. Let
${\bf P}(t) = {\bf P}_\mathcal{K}(t)\oplus{\rm P}_\triangle(t)$ give the probability that the system is in a state $i\in\mathcal{K}\cup\triangle$ at time $t$; the continuous time limit yields
\begin{equation}
  \dot{\bf P}(t) = {\bf P}(t)\cdot{\bf Q}
  \quad\Rightarrow{\bf P}(t)={\bf P}(0)\cdot\exp({\bf Q}t)\label{CTMC}.
\end{equation}
The absorbing transition matrix ${\bf Q}$, illustrated in figure \ref{rmfig}, has a structure
\begin{equation}
  {\bf Q} = \begin{bmatrix}
            {\bf Q}_\mathcal{K}  & {\bf k}^u\\
            {\bf 0}^{\rm T} & 0
          \end{bmatrix},
\end{equation}
where $({\rm Q}_\mathcal{K})_{ij}\equiv k_{ij}-k^{\rm tot}_i\delta_{ij}$ for $i,j\in\mathcal{K}$, ${\bf 0}$ is a vector of zeros and $({\bf k}^u)_i\equiv k^{\rm un}_i$. From the structure of $\bf Q$ one finds that
\begin{equation}
{\bf P}_\mathcal{K}(t) = {\bf P}_\mathcal{K}(0)\cdot\exp({\bf Q}_\mathcal{K}t).
\end{equation}
As the probability of transition to $\triangle$ from a state $i$ at a time $t$ is given by $[{\bf P}_\mathcal{K}(t)]_ik^{\rm un}_i$, the expected residence time is simply
\begin{equation}
  \langle\tau^{res}\rangle =
  \int_0^\infty t{\bf P}_\mathcal{K}(t)\cdot{\bf k}^u{\rm d}t =  \frac{
  {\bf P}_\mathcal{K}(0)\cdot{\bf Q}^{-2}_\mathcal{K}\cdot{\bf k}^u
  }{
  {\bf P}_\mathcal{K}(0)\cdot{\bf Q}^{-1}_\mathcal{K}\cdot{\bf k}^u
  }.
\end{equation}
Defining a vector of ones ${\bf 1}_\mathcal{K}$, it is simple to show that ${{\bf Q}_\mathcal{K}}\cdot{\bf 1}_{\mathcal{K}} = -{\bf k}^u$, giving the further simplification
\begin{equation}
  \langle\tau^{res}\rangle = -{\bf P}_\mathcal{K}(0)\cdot{\bf Q}_\mathcal{K}^{-1}\cdot{\bf 1}_\mathcal{K}. \label{res_t}
\end{equation}
This expression for the residence time can be evaluated by solving the linear equation ${\bf Q}^{\rm T}_\mathcal{K}\cdot{\bf x}={\bf P}_\mathcal{K}(0)$ to give $\langle\tau^{res}\rangle = -{\bf x}\cdot{\bf 1}_\mathcal{K}$.


Since $\langle\tau^{res}\rangle$ quantifies the quality of the current CTMC, it is natural to use it as an objective function guide further improvement given a computational effort $\delta c$ that can be invested.
To best harness massively parallel computational resources, the optimal allocation will be expressed as an allocation distribution $\{s_i\}$ which gives the proportion of workers assigned to each state $i\in\mathcal{K}$. The computational effort $c_i$ allocated to state $i$ is therefore
\begin{equation}
  \delta c_i \equiv s_i\delta c,\quad \sum_{i\in\mathcal{K}} s_i \equiv 1.
\end{equation}
In the expression (\ref{res_t}) for the residence time, only the unknown rates are affected by the additional computational work, giving to leading order in $\delta c$ a change in the residence time of (see appendix \ref{app:dme})
\begin{equation}
  \frac{\delta\langle\tau^{res}\rangle}{\delta c} = -\sum_{i\in\mathcal{K}} s_i \frac{\delta k^{\rm un}_i(\beta_{\rm L})}{\delta c_i}
  \left[{\bf P}_\mathcal{K}(0)\cdot{\bf Q}_\mathcal{K}^{-1}\right]_i
  \left[{\bf Q}_\mathcal{K}^{-1}\cdot{\bf 1}_\mathcal{K}\right]_i
  \label{dtau}
\end{equation}
where $-\delta k^{\rm un}_i/{\delta c_i}$ is precisely the maximized statewise cost function (\ref{obj_fun}) found in the previous section, evaluated at its maximum, i.e. at the high temperature $\beta_H$ which maximizes $-\delta k^{\rm un}_i/{\delta c_i}$. As equation (\ref{dtau}) takes the form of an inner product the optimal choice of $s_i$ is simply
\begin{equation}
  s_i = \eta
  \frac{\delta k^{\rm un}_i(\beta_{\rm L})}{\delta c_i}
  \left[{\bf P}_\mathcal{K}(0)\cdot{\bf Q}_\mathcal{K}^{-1}\right]_i
  \left[{\bf Q}_\mathcal{K}^{-1}\cdot{\bf 1}_\mathcal{K}\right]_i,\label{opt_alloc}
\end{equation}
where $\eta^{-1} = \sum_{j\in\mathcal{K}}s_j$ ensures normalization. Solving the linear equation $\left[{\bf Q}_\mathcal{K}\right]\cdot{\bf y}={\bf 1}_\mathcal{K}$, one gets $s_i = \eta{\rm x}_i{\rm y}_i({\delta k^{\rm un}_i}/{\delta c_i})$. This simple procedure insures that additional resources are optimally invested in order to maximize $\langle\tau^{res}\rangle$ at the smallest computational cost. In practice, the optimal allocation is periodically updated using the latest CTMC.

The optimal allocation has a clear interpretation using two expressions that follow from equation (\ref{res_t}) for the residence time. As the inner product with ${\bf 1}_\mathcal{K}$ is simply a sum over the known states, the second term in (\ref{opt_alloc}), $-\left[{\bf P}_\mathcal{K}(0)\cdot{\bf Q}_\mathcal{K}^{-1}\right]_i$, is simply the expected time spent in a state $i$ conditional on the initial distribution ${\bf P}_\mathcal{K}(0)$, which when summed over all states yields $\langle\tau^{res}\rangle$. If we instead take the initial distribution to be a delta function on a state $i$, the third term in (\ref{opt_alloc}), $-\left[{\bf Q}_\mathcal{K}^{-1}\cdot{\bf 1}_\mathcal{K}\right]_i$, can be interpreted as the expected residence time in the known network, conditional on starting from a state $i$.
The allocation of computational work to a state is thus a product of three factors- the degree to which the unknown rate will change under additional sampling,
the amount of time (on average) spent in the state before absorption under the desired initial conditions, and the characteristic residence time of trajectories starting in the state. If a state is very well sampled, the last two factors might be large, but the change in the unknown rate with additional sampling will be very small, suppressing the allocation weight. Conversely, a poorly sampled state might be rarely visited and have a small residence time, but the change in the unknown rate will be very large, increasing the allocation weight.
In this manner, TAMMBER is able to allocate computational work to a state according to a {\em global} measure of the state's influence on the ensemble of trajectories in the known state space, dependent only on the prescribed the initial condition ${\bf P}_\mathcal{K}(0)$.

\section{TAMMBER Simulation Code}\label{secTAM}
We have implemented the TAMMBER workflow, illustrated in figure \ref{rmfig}, within the ParSplice\cite{perez2015} simulation code, which provides the underlying framework for generating state-to-state trajectories, state identification and asynchronous control over the requested work using massively parallel computational resources. MD trajectories themselves are generated by the LAMMPS molecular dynamics package\cite{LAMMPS}; after a 1ps thermalization and dephasing stage (which is repeated if a transition occurs\cite{perez2015}), a snapshot of the system is recorded 2-4 times over each ps trajectory segment, with the final snapshot relaxed and analyzed\cite{perez2015} to check for transitions between metastable states. If a transition is detected, the intermediate snapshots are relaxed and analyzed to find a more precise transition time and to check for multiple transitions, which can occur if a low barrier is found at a high temperature. Transition times and pathways are sent back to the central task manager, with new transitions submitted for a climbing-image NEB calculations\cite{neb} and, if desired, a Hessian prefactor calculations using LAMMPS force calls and the FIRE minimization routine\cite{FIRE}.

The central task manager of TAMMBER analyzes, at regular intervals, all of the state-to-state trajectory data using the multi-temperature TAD formalism outlined in section \ref{secTAD} to produce a list of time ordered first passage times and final states for each state.
The dynamical data $\{\tau_{ij}\}$ and static data $\nu^0_{ij},E_{ij}$ for each transition is then used to produce an estimate of the rate prefactor using the Bayesian estimators derived in section \ref{secRATES}. With knowledge of the individual transition rates $k_{ij}(\beta) = \nu_{ij}\exp(-\beta\Delta E_{ij})$ at the desired temperature, we can estimate
$\langle k_u(\beta)\rangle$ and $\langle k^2_u(\beta)\rangle$ using the Bayesian posterior distribution for the total escape rate (\ref{posterior_tk})
and therefore fully populate the matrix $\bf Q$ for the absorbing Markov chain (\ref{CTMC}) at the low temperature $\beta_{\rm L}$.
The quality of this CTMC is assessed by computing $\langle\tau^{res}\rangle$ for a given initial distribution and further allocation of resources carried out according to the distribution (\ref{opt_alloc}) that maximizes the rate of increase of $\langle\tau^{res}\rangle$. The cycle then repeats until $\langle\tau^{res}\rangle$ is deemed sufficiently small, or computational resources are exhausted. In the next section, we first test TAMMBER against an exactly known total rate matrix using kinetic Monte Carlo to generate trajectories, then use TAMMBER to explore the evolution of interstitial clusters in iron.


\subsection{Validation using a known rate matrix}
A key component of the TAMMBER code is to estimate $\langle k^{\rm un}_i \rangle$, the unknown (or remaining) rates from each explored state, in order to construct an absorbing CTMC which both allocates resources and provides a metric for the degree of exploration. To validate our estimator for $\langle k^{\rm un}_i \rangle$, we replaced the molecular dynamics engine with a simple kinetic Monte Carlo (kMC) routine\cite{bortz1975} using a prescribed matrix rate matrix $k_{ij} = \nu_{ij}\exp(-\beta\Delta E_{ij})$ constructed at any temperature from a pre-specified list of energy barriers $\Delta E_{ij}$ and prefactors $\nu_{ij}$.
To ensure the rate matrix satisfies detailed balance, we assign a free energy $F_i = E_i -\beta^{-1}\log\omega_i$ to each state and a symmetric saddle point free energy $F_{ij} = F_{ji} = E_{ij} -\beta^{-1}\log\omega_{ij}$, then build barriers and prefactors through $\Delta E_{ij} = E_{ij} - E_i$ and $\nu_{ij} = \omega_{ij}/\omega_i$. The energies were drawn from a uniform distribution and prefactors from a log uniform distribution between 0.01THz and 100THz.

When using the kMC backend, we have access to the exact remaining rate at any point in the simulation, which can be compared to our estimates $\langle k^{\rm un}_i \rangle$. Figure \ref{figure3} demonstrates the estimate of the unknown rate for a single state against the simulated computational cost (performing MD, identifying states and NEB calculations) at a range of fixed TAD temperatures $\beta^{-1}_{\rm H}$, and the TAMMBER process, which uses a variable TAD temperature determined by maximizing the benefit function $-\delta\langle k^{\rm un}_i \rangle/\delta c_i$, equation (\ref{obj_fun}). It can be seen that TAMMBER successfully adjusts the TAD temperature to decrease the unknown rate as fast as possible with computational effort, whilst the estimate $\langle k^{\rm un}_i \rangle$ decreases with increasing sampling time. Importantly, the estimated unknown rate is greater than the actual remaining rate, meaning that we can have high confidence that the predicted residence times are conservative. This behavior emerges naturally from our Bayesian estimator; given only the knowledge that rare events are Poisson random variables (through the likelihood function) our estimate for the remaining rate cannot be significantly lower than the inverse time spent in the state, i.e., one cannot exclude the possibility of a given $k^{\rm un}_i$ remaining without running dynamics for a time of order $1/k^{\rm un}_i$. Whilst it is in principle possible to improve the estimator by encoding knowledge of the rate distribution into a Bayesian prior, such information is typically not available in atomistic simulation, so the estimator (\ref{posterior_tk}) is a good choice.

We have also used the kMC backend to test self-optimizing capability of TAMMBER beyond a single state. Two rate matrices were generated, each with 100 states and on average 40 connections per state, but with a different distribution of energy barriers. We chose a high connectivity to ensure each state has a similar spectrum of escape rates, whilst as before the target temperature was 300K and TAD temperatures between 300K and 1500K were considered. To investigate the response of our control protocol \ref{obj_fun}, the first rate matrix (System 1) had barriers drawn between 0.25eV and 1eV, whilst the second rate matrix (System 2) had barriers drawn between 0.5eV and 1.25eV, suggesting a higher optimal temperature. As can be seen in \ref{figure3b}, TAMMBER is able to self-optimize for these two systems; the mean optimal TAD temperature for System 1 is around 600K, whilst for System 2 this rises to 1200K.

For the example cases considered here, where each state has a similar spectrum of escape rates, the spread of optimal temperatures across the states is relatively narrow, but in a general case this can vary significantly as a function of rate spectrum and time spent in the state. In general, the optimal temperature will start at the lowest value, quickly rise as state time is accumulated before the first event is observed, then fall to a degree dependent on the discovered transition rates.

\begin{figure}
  \includegraphics[width=\columnwidth]{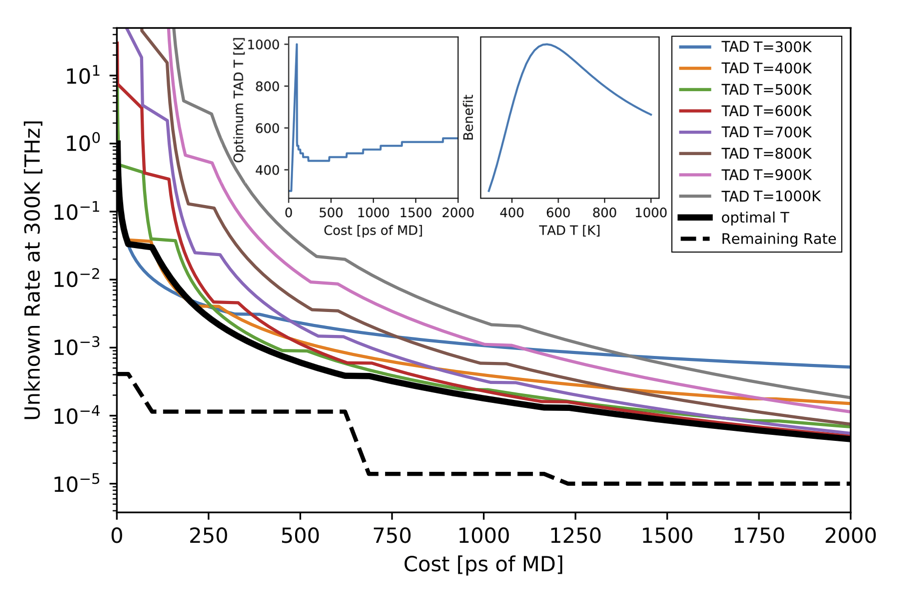}
  \caption{Comparison of TAMMBER and typical TAD sampling a single state at a target temperature of 300K, using kMC to generate escape times, with an estimated computational cost in units of ps of MD. The optimal TAD scheme implemented in TAMMBER is able to find the optimum instantaneous TAD temperature to reduce the unknown rate for minimal computational cost, and thus is able to autonomously outperform constant temperature TAD. Left inset: optimal temperature during simulation. Right inset: The objective function (\ref{obj_fun}) at the end of the simulation. \label{figure3}}
\end{figure}
\begin{figure}
  \includegraphics[width=\columnwidth]{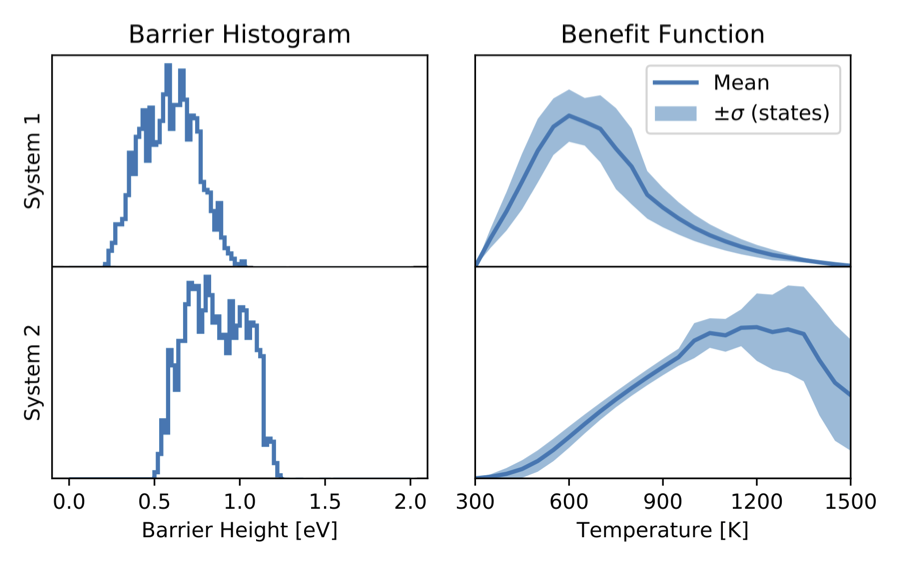}
  \caption{Self-optimization of the TAMMBER code for two test systems. Left: Histogram of energy barriers. Right: Mean and standard deviation of the benefit function across the range of temperatures. System 2 has a systematically larger barrier spectrum than system 1, leading to an increase in the optimal TAD temperature.\label{figure3b}}
\end{figure}

\subsection{Interstitial capture by C15 clusters in Iron}
As a preliminary application of TAMMBER, we have investigated the capture of mono-interstitial dumbbell defects\cite{fu2005} by C15 tetra-interstitial clusters\cite{marinica2012} using an embedded atom potential model of iron\cite{malerba2010b}. C15 clusters have been observed in irradiation damage simulations\cite{zarkadoula2013} and are known to be the most stable interstitial arrangement for small defect sizes\cite{Dezerald2014C15}, but their connection to the wider energy landscape of an irradiated material is still largely unexplored. In particular, C15 defects have been observed to act as sinks for mono-interstitials, resulting in C15 growth which is assumed to play an important role in the evolution of the defect population\cite{zhang2015formation}. However, due to the vast energy landscape of a defective material quantitative statements on the nature of this capture processes, beyond observation of individual trajectories, is very challenging to calculate by traditional methods.

In our simulations, a C15 structure\cite{marinica2012} was formed from 4 interstitial atoms in a 10x10x10 cubic supercell before adding a further interstitial atom nearby, forming a dumbbell under further relaxation. Minimizing the hydrostatic pressure changed the final energy by less that 0.01 eV, consistent with the known small formation volume of these defects\cite{marinica2012}. The final system, illustrated in figure \ref{figure5}A, contained 2005 atoms. TAMMBER performed constant volume TAD MD simulations using an underdamped Langevin thermostat\cite{LAMMPS}, with a target temperature of 300K and possible TAD temperatures between 400K and 900K. Resource allocation was determined using the scheme detailed above, with the initial distribution being a delta function $ [ {\bf P}_\mathcal{K}(0) ]_i = \delta_{ij} $ on the starting state of a separated dumbbell and C15 tetra-interstitial. The upper temperature threshold is limited by the presence of significant anharmonic effects on the transition rate which violate the harmonic approximation used in TAD; efficient anharmonic rate theory implementations\cite{swinburne2017} would therefore be extremely beneficial to further extend the range of TAD temperatures that can be used. As anharmonic vibrational effects typically act to increase transition rates, it can be shown that the inclusion of anharmonic effects would act to increase the expected residence time of the observed network and thus our present results can be considered a lower bound.

After 12 hours of operation on 2160 processors, TAMMBER had identified 2664 metastable states with 7676 connecting barriers from around 2$\mu$s of high temperature MD. The expected residence time conditional on ${\bf P}_\mathcal{K}(0)$ in the set of known states was found to be 43.4 seconds at 300K, a testament to the timescales that can be accessed by the massively parallel temperature accelerated dynamics controlled by TAMMBER.

\begin{figure}[!ht]
  \includegraphics[width=\columnwidth]{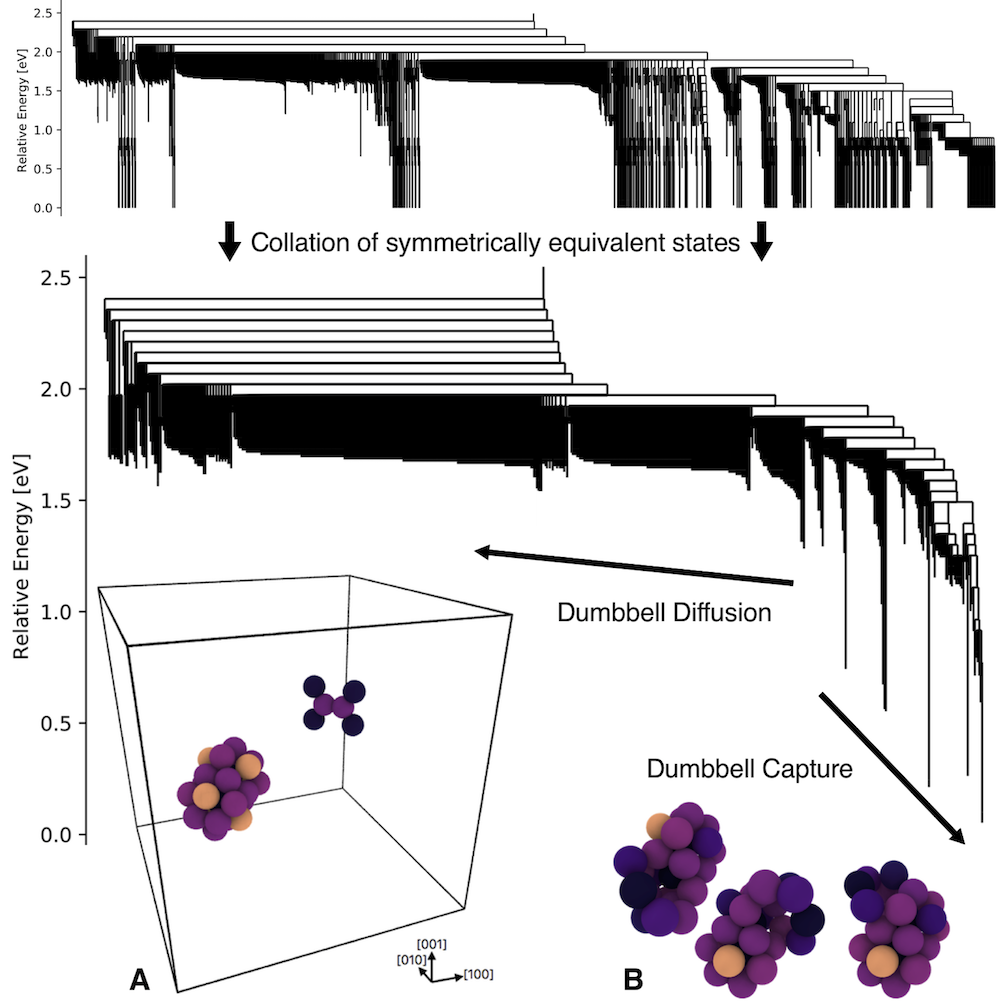}
  \caption{Disconnectivity graph\cite{wales_energy_2003} for states found by TAMMBER with the tetra-C15 and dumbbell system studied shown in inset A. As discussed in the main text, grouping symmetrically equivalent states leads to a significant reduction in the number of states and simplifies the graph structure. Whilst all the dumbell capture states (inset B) reside in the superbasin, distinct dumbell states also exist inside this superbasin at relatively low energies, meaning this illustration is not a perfect representation of the coarse grained landscape. Nevertheless, the Markov chain analysis shows the system remains in a penta-C15 states for multiple seconds at 300K.}
  \label{figure5}
\end{figure}

The energy landscape, illustrated through a disconnectivity graph\cite{wales_energy_2003} in figure \ref{figure5}, consists of a large number of states corresponding to dumbbell diffusion (shown in figure \ref{figure5}A), with a smaller number of low energy states corresponding to dumbbell capture (figure \ref{figure5}B). As discussed in the next section, whilst the clear superbasin shown in figure \ref{figure5} contains all the dumbbell capture states, distinct dumbbell diffusion states also exist at relatively low energies, meaning the superbasin structure is an illustrative but imperfect representation of the coarse grained landscape. Due to the high stability of the C15 tetra-interstitial, the number of states is expected to scale only linearly with the size of the system, resulting in the relatively low number of states found in this example.

Figure \ref{figure6} gives a more detailed presentation of the final state of the TAMMBER simulation. Figure \ref{figure6}A shows the wide distribution of energy barriers found, demonstrating the need for an adaptive parametrization to optimally sample the highly heterogeneous energy landscape. The peak around 0.26eV corresponds to dumbbell migration, with higher energy barriers typically corresponding to escape pathways from the superbasin of captured states. Figure \ref{figure6}B show the relative state energies, with a peak at the minimum energy for the superbasin of capture states, whilst the large higher energy peak is for dumbbell migration states. Figure \ref{figure6}C shows the effective low temperature TAD time versus high temperature MD time for each state. The slope on double log plot is equal to the temperature ratio $\beta_{\rm L}/\beta_{\rm H}$, as can be seen from equations \ref{emin} and \ref{state_time}. The deepest states have the highest optimum TAD temperature, and can clearly be seen as the upper envelope to the the scatter low temperature TAD times, whilst the lowest envelope is simply an equality ($\beta_{\rm L}=\beta_{\rm H}$). The scatter in slope is a demonstration of the range of optimal temperatures throughout the run; deep capture states were typically sampled at 900K, whilst the dumbbell diffusion states were typically sampled at around 550K. Finally, figure \ref{figure6}D shows a histogram of the low temperature unknown rates $k^{\rm un}_i(\beta_{\rm L})$. States which are not deemed influential to the overall behavior by the Markov chain analysis receive little to no sampling and thus possess a high unknown rate, leading to the significant upper peak.

At the end of the initial TAMMBER simulation, it was observed that the resultant Markov chain predicts a long residence time in the superbasin of low lying `dumbbell capture` states (shown in figure \ref{figure5}B). To further explore this superbasin, TAMMBER was restarted using the previously generated trajectory and transition barrier information and ran for a further 4 hours on 2160 cores with a new initial distribution, namely a delta function on the best sampled dumbbell capture state, giving an expected residence time of $\langle\tau_{\rm res}\rangle=57.6s$ at 300K, with 21 states having an expected visit time of more than 0.1 seconds, from a total effective low temperature time of $\sum_i\tau_i(\beta_{\rm L})=2.98\times10^4$s. This scale separation between the total low temperature time and the residence time is a consequence of the structure of the energy landscape; as superbasin states are frequently revisited, the unknown rate must be significantly lower than the total known escape rate to to ensure long trajectories before absorption. AMD techniques are thus essential to provide efficient sampling of the energy landscape, as otherwise the the raw sampling MD time greatly exceeds the typical residence time of the found transition network\cite{bhoutekar2017new}.

Upon a detailed investigation of the observed system configurations, it was found that a significant number of states were identical to each other up to a reindexing of atoms or an operation of the crystal's symmetry group. Exploitation of these symmetries are clearly highly desirable, as the high temperature MD trajectories and found escape times across all identical states to be collated, resulting in more efficient sampling, smaller unknown rates and a more compact description of the transition network. As the effective state time is increases to the power of the temperature ratio used in TAD, consolidation of MD sampling can produce very large decreases in the unknown rates.
Identification of symmetrically equivalent states is possible using graph isomorphism algorithms\cite{NAUTY} on the connectivity graphs used to identify states in TAMMBER. Using the graph isomorphism algorithm to construct a map to the reduced set of symmetrically inequivalent states, we reprocessed the TAMMBER simulation output to construct new effective state times, transition rates and unknown rates to build a new Markov chain in the symmetrically reduced state space. We find a new transition network of 626 states, illustrated in figure \ref{figure6}. The new residence time with a delta function on the same lowest energy superbasin state is now $\sum_i\tau_i(\beta_{\rm L})=7.38\times10^6$s with a residence time of 80.9 seconds. This very large difference between the total state and validity is due to the high degree of degeneracy (318 states) of the lowest lying dumbbell capture state, resulting in an excessively long effective state time of $4.2\times10^6$s, which would not be allocated in a symmetry-aware resource management scheme. The development of such a scheme in TAMMBER is clearly highly desirable but raises a number of subtle issues which are beyond the scope of the present paper. In our final section, we use the symmetrically reduced Markov Chain developed above to investigate superbasin escape times and explore the consequences of possessing, through the unknown rates, uncertainty quantification on the completeness of the discovered network.

\begin{figure}[!ht]
  \includegraphics[width=\columnwidth]{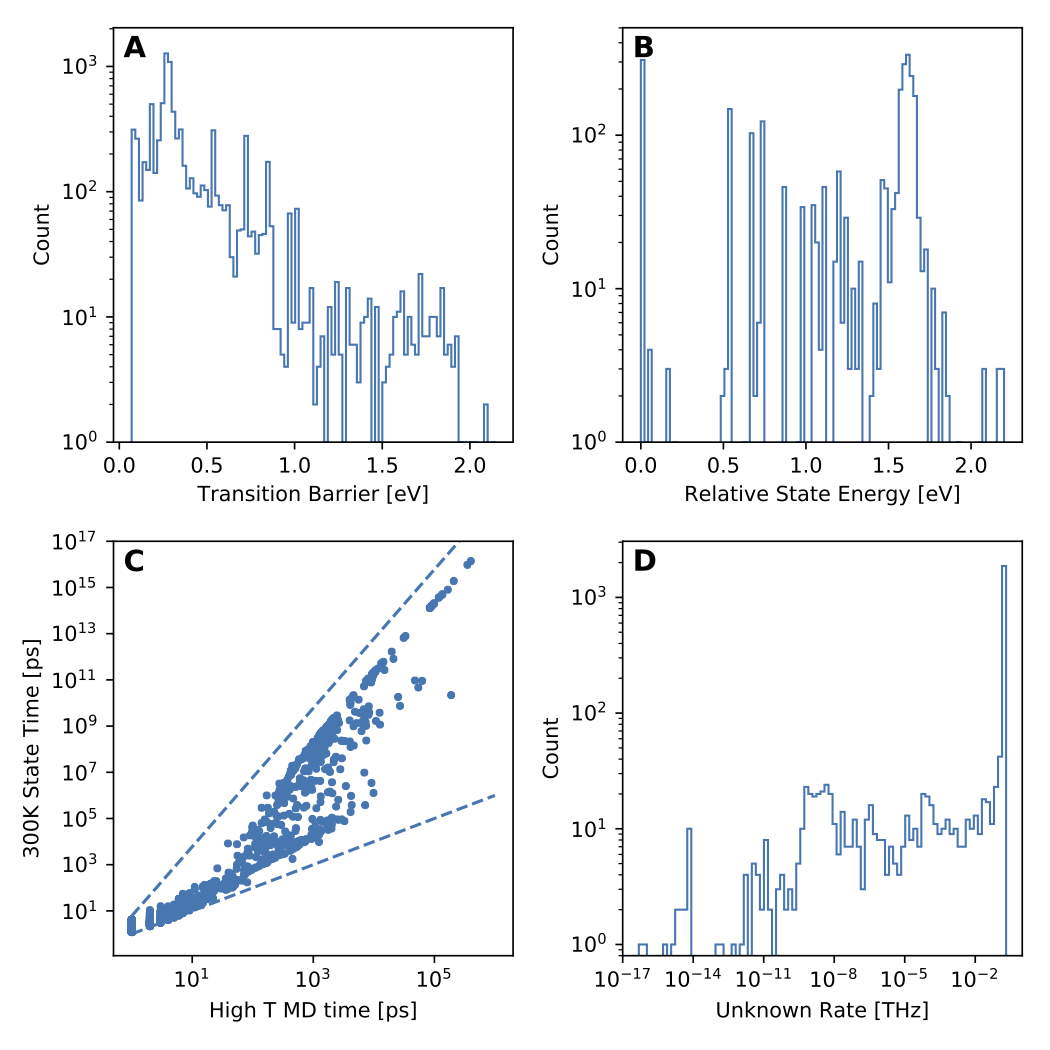}
  \caption{Summary of the TAMMBER simulation for the C15-dumbbell system, discussed in the main text. A: Histogram of energy barriers. B: Histogram of state energies. C: Effective low temperature time versus high temperature MD time. D: Histogram of unknown rates}
  \label{figure6}
\end{figure}

\section{Discussion: uncertainty quantification of transition network observables}\label{secCON}
The central goal of the present paper was to construct, with rigorous uncertainty quantification, a transition network from atomistic simulations with a maximally long residence time in the found state space. The previous section demonstrated that extremely long residence times are readily accessible using our method. In this final section we provide a preliminary exploitation of the discovered transition network, in particular accounting for of uncertainty quantification provided by the unknown absorbing rates. A full exploration of these ideas, and a detailed examination of their use when transitioning to higher scale simulation scheme such as kinetic Monte Carlo, will be the subject of future work.
A natural observable to extract from the transition network is the expected escape time from the dumbbell capture superbasin. This is clearly an important input for coarse grained models of interstitial cluster evolution, informing the degree to which C15 clusters can be considered as pure sinks for mono-interstitial defects, which can otherwise collate into highly mobile prismatic dislocation loops.
To calculate a superbasin escape time, we ask for the first escape time from a collection of states $\mathcal{A}$, here the lowest energy dumbbell capture states (figure \ref{figure6}B). This can simply be achieved by artificially making all the remaining states $\mathcal{B}=\mathcal{K}\setminus\mathcal{A}$ an absorbing set. Similar ideas are regularly employed in the biochemical community\cite{wales2002discrete}, though the inclusion of an unknown rate to account for sampling incompleteness is novel to the best of our knowledge.
Defining the known rate matrix on $\mathcal{A}$ as ${\bf Q}_{\mathcal{A}}$ one can then define {\it two} sets of absorbing rates, namely the previously estimated unknown rates from $\mathcal{A}$ to $\triangle$ and the sum of all rates from $\mathcal{A}$ to $\mathcal{B}$:
\begin{equation}
  \left[{\bf k}^u_{\triangle}\right]_i={\rm k}^u_i,\quad
  \left[{\bf k}^u_{\mathcal{B}}\right]_i =\sum_{j\in\mathcal{B}} {\rm k}_{ij},\quad
  i\in\mathcal{A}.
\end{equation}
Restricting the initial distribution of states ${\bf P}_\mathcal{A}(0)$ to some distribution over $\mathcal{A}$, one can define a very useful convergence measure for averages over trajectories from $\mathcal{A}$ to $\mathcal{B}$, namely the probability of absorbing to $\mathcal{B}$ instead of $\triangle$, given by
\begin{equation}
  {\rm P}_{\mathcal{B}<\triangle} = {\bf P}_\mathcal{A}(0)\cdot{\bf Q}^{-1}_{\mathcal{A}}\cdot{\bf k}^u_{\mathcal{B}},\quad \lim_{{\rm k}^{\rm u}_\triangle\to0}{\rm P}_{\mathcal{B}<\triangle}=1,
\end{equation}
where the final limit corresponds to convergence to the complete model. The expected first passage time from $\mathcal{A}$ to $\mathcal{B}$, conditional on not absorbing to $\triangle$, reads
\begin{equation}
  \langle\tau^{abs}_{\mathcal{A}\to\mathcal{B}}\rangle={\bf P}_\mathcal{A}(0)\cdot{\bf Q}^{-2}_{\mathcal{A}}\cdot{\bf k}^u_{\mathcal{B}} / {\rm P}_{\mathcal{B}<\triangle}.\label{abs_t}
\end{equation}
However, when ${\rm P}_{\mathcal{B}<\triangle}$ is small, absorption to $\triangle$ is much more likely and thus the true first passage time from $\mathcal{A}$ to $\mathcal{B}$ is expected to be much greater than the current residence time, meaning (\ref{abs_t}) is likely to be a significant underestimate. One possible strategy to investigate the dependence of $\langle\tau^{abs}_{\mathcal{A}\to\mathcal{B}}\rangle$ on sampling incompleteness is to
`artificially' take the limit perfect sampling limit
\begin{equation}
  \langle\tau^{abs}_{\mathcal{A}\to\mathcal{B}}|{\rm k}^{\rm u}_\triangle=0\rangle \equiv \lim_{{\rm k}^{\rm u}_\triangle\to0} \langle\tau^{abs}_{\mathcal{A}\to\mathcal{B}}\rangle,
\end{equation}
corresponding to the prediction of approaches without any uncertainty quantification. Another approach is to recognize that the conditional expectation in (\ref{abs_t}) is biased by sampling only from the subset of trajectories that absorb to $\mathcal{B}$ before $\triangle$. An approximate form for the unbiased first passage time $\langle\tau^{abs}_{\mathcal{A}\to\mathcal{B}}\rangle_\infty$ can be obtained by assuming absorption from $\mathcal{A}$ to $\mathcal{B}$ or $\triangle$ are two first order Poisson processes with mean times $\langle\tau^{abs}_{\mathcal{A}\to\mathcal{B}}\rangle_\infty$ and $\langle\tau_{\rm res}\rangle$; it is simple to show that this gives the approximate expression for the unbiased first passage time of
\begin{equation}
  \langle\tau^{abs}_{\mathcal{A}\to\mathcal{B}}\rangle_\infty = \left(\frac{1}{{\rm P}_{\mathcal{B}<\triangle}}-1\right)\langle\tau_{\rm res}\rangle.
\end{equation}

Coming back to the case of the C15 defects, Whilst the disconnectivity graph has a clear superbasin structure, a detailed inspection shows that a number of lower energy states have a distinct, separate dumbbell structure, meaning that the `true' capture superbasin has a more complex structure than that implied by the illustration in figure \ref{figure5}. The set $\mathcal{A}$ of capture states were thus chosen to be the minimal set of connecting states to the found global minimum where a distinct dumbbell structure could not be found, consisting of 63 symmetrically inequivalent states, or around 500 states of the original network.

In figure \ref{figure7}B, we plot ${\rm P}_{\mathcal{B}<\triangle}$ along with the residence time $\langle\tau_{\rm res}\rangle$, the conditional first passage time $\langle\tau^{abs}_{\mathcal{A}\to\mathcal{B}}\rangle$ and corrected first passage times $\langle\tau^{abs}_{\mathcal{A}\to\mathcal{B}}|{\rm k}^{\rm u}_\triangle=0\rangle$ and  $\langle\tau^{abs}_{\mathcal{A}\to\mathcal{B}}\rangle_\infty$ across for temperatures from the low target temperature 300K to 900K, the highest temperature considered in our simulations. It can be seen that at low temperatures ${\rm P}_{\mathcal{B}<\triangle}$ is very small, leading the conditional first passage time to converge to the network residence time, indicating that the current quality of the network in insufficient to "certify" that the predicted times are correct, even if
the corrected times are essentially in perfect agreement with each other. In other words, the model cannot be used to exclude the possibility that other, yet undiscovered mechanisms, could affect the predicted times at low temperatures.
In contrast, at higher temperatures ${\rm P}_{\mathcal{B}<\triangle}\to1$, the residence time exceeds the first passage time, with all estimates for the first passage time converging. From the Arrhenius gradient an effective energy barrier for the superbasin escape of 1.41eV is found, which closely corresponds the escape process illustrated in figure \ref{figure7}A.

Whilst the network produces a reliable superbasin escape time at high temperature, the large differences between the residence time and the corrected first passage times at low temperature, or equivalently the small values of ${\rm P}_{\mathcal{B}<\triangle}$, demonstrate that care must be taken when constructing transition networks from atomistic simulations. The objective function (\ref{obj_fun}) used in the current work was focussed on optimizing a particular measure of transition network quality, the expected residence time. In future work, we will further develop the approach presented here to specifically address the issue of converging more targeted quantities such as superbasin escape times.

\begin{figure}[!ht]
  \includegraphics[width=\columnwidth]{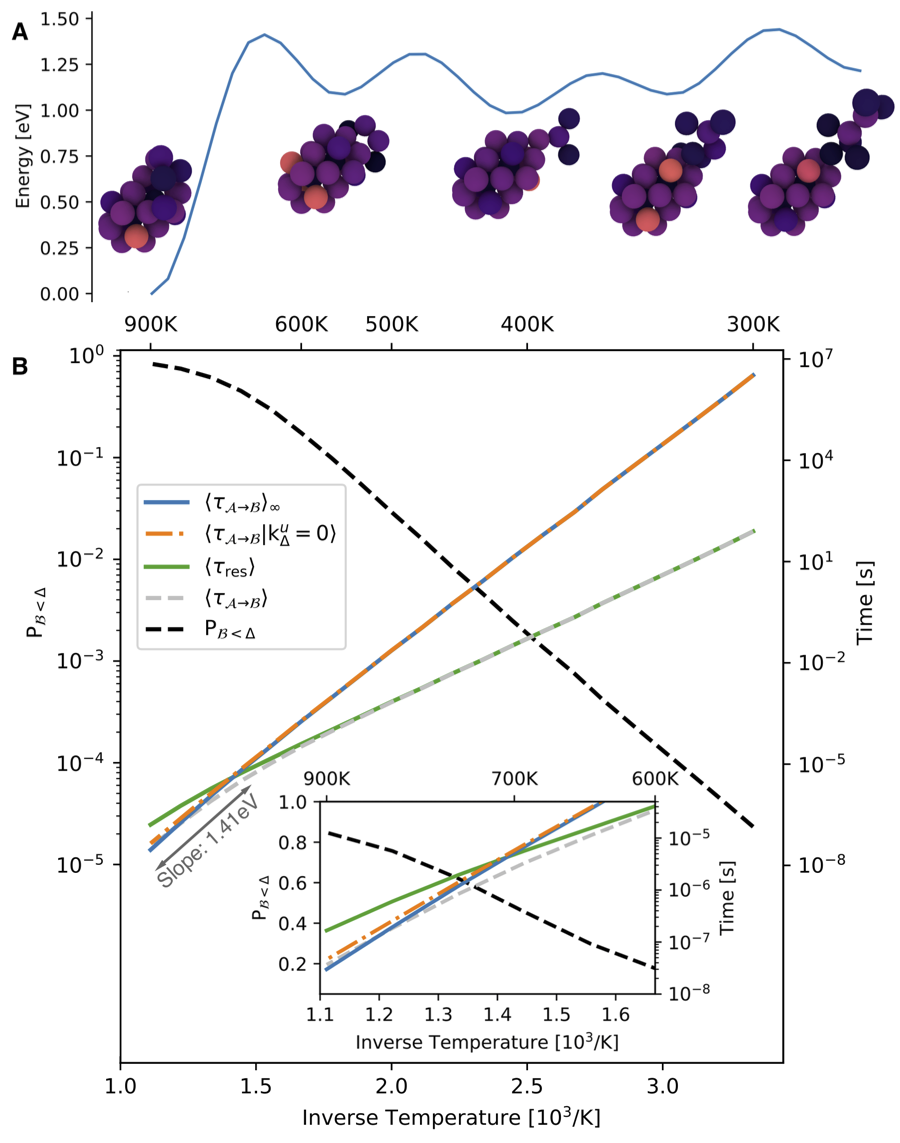}
  \caption{Analysis escape from the dumbbell capture superbasin tetra-C15 and shown in \ref{figure6}B. Below: calculated residence times and unbiased first passage times across a range of temperatures. As discussed in the main text, the first passage time estimates all converge when the probability of escape before absorption is high. Above: the minimum energy path for superbasin escape. The highest saddle point energy of 1.44eV agrees well with the found Arrhenius slope of 1.41 eV.}
  \label{figure7}
\end{figure}

\section{Conclusions}
In this paper we have introduced a method to generate large networks of transition rates from atomistic simulations, sampling the energy landscape with a novel form of self-optimizing temperature accelerated dynamics. Bayesian estimators were developed that quantify sampling incompleteness in the form of an absorbing unknown rate for each system state. Due to sampling incompleteness, trajectories in the observed rate network have a finite lifetime before absorption.
The presented method, TAMMBER, determines the optimal allocation of computational resources `on-the-fly` in order to find new states and transition pathways, with the goal of maximize the expected time in the known transition network before absorption, conditional on a user specified initial condition. After validation on exactly known transition networks TAMMBER was applied to the capture of interstitials by C15 clusters in an EAM model of Iron, reaching expected absorption times of more than 80s at 300K. It was found that sampling completeness could be considerably improved by consolidating symmetrically equivalent states; incorporation of symmetry considerations into the TAMMBER allocation scheme is an immediate topic for future work.

The transition network was then used to explore superbasin escape times, with expressions derived for the average escape rate in terms of the absorbing rates. The uncertainty quantification indicated that whilst converged results can be produced when the predicted escape time is less than the network residence time,
building statistical confidence on long-time, low-temperature, behavior proves extremely challenging, as results can be strongly affected
by the degree of sampling completeness, an observation which is likely to be widely applicable across many coarse grained modeling approaches. The further development of
optimal strategies such as this one to reduce the often surprisingly large uncertainty sensitivity are therefore urgently needed.

This material is based upon work supported by the U. S. Department of Energy, Office of Nuclear Energy and Office of Science, Office of Advanced Scientific Computing Research through the Scientific Discovery through Advanced Computing (SciDAC) project on Fission Gas Behavior and used computing resources provided by the Los Alamos National Laboratory Institutional Computing Program. Los Alamos National Laboratory is operated by Los Alamos National Security, LLC, for the National Nuclear Security administration of the U.S. DOE under Contract No. DE-AC52-06NA25396.

\appendix

\section{Moments of the posterior distribution}\label{app:anal}
For $N$ seen transitions, we define $a_j \equiv k^{\rm obs}_i-k^{\rm obs}_{i;j}$, $j\in[0,N_i-1]$. As $k^{\rm obs}_{i;0}=0$ (no observed rate before the first event), equation (\ref{posterior_tk}) for the posterior distribution can then be written
\begin{align}
  \pi(k^{\rm un})&=e^{-k^{\rm un}\tau_i}\prod_{j=1}^{N-1}\left(k^{\rm un}+a_j\right)\\
                    &=e^{-k^{\rm un}\tau_i}\sum_{r=0}^{N-2} \left(k^{\rm un}\right)^r A_r,
\end{align}
where $A_r$ is the sum of all $^{N-2}C_r$ possible combinations of $r$ elements from $\{a_j\}_1^{N-1}$. By considering the change in $A_r$ when expanding the number of terms in the product, the $A_r$ can be evaluated by a simple recursion. Using the integral relation $\int_0^\infty k^n e^{-kt}{\rm d}k = n! t^{-(n+1)}$ we can thus write
\begin{equation}
  \langle \left( k^{\rm un}_i\right)^n \rangle = \frac{\sum_{r=0}^{N-2} (r+n)! A_r  \tau_i^{-r}} {\tau_i^n\sum_{r=0}^{N-2} r! A_r \tau_i^{-r}}.
\end{equation}

\section{Derivative of an inverse matrix element}\label{app:dme}
Consider the known derivative $\partial_l {\rm A}_{ij}$ of an element of a matrix $\bf A$. To calculate the derivative of the inverse matrix element $\partial_l ({\rm A}^{-1})_{ij}$, we apply the chain rule to the trivial result $\partial_l \left({\bf A}\cdot{\bf A}^{-1}\right)\equiv 0$ then premultiply by ${\bf A}^{-1}$ to obtain
\begin{equation}
  \partial_l {\bf A}^{-1} = - {\bf A}^{-1}\cdot \partial_l {\bf A} \cdot {\bf A}^{-1}.
\end{equation}
Note the nontrivial ordering of the matrix product. From the structure of the known rate matrix ${\bf Q}_\mathcal{K}$ we have
\begin{equation}
  \frac{\partial}{\partial k^{\rm un}_l} \left[{\bf Q}_\mathcal{K}\right]_{ij} = -\delta_{ij}\delta_{il}.
\end{equation}
This gives the inverse derivative as
\begin{equation}
  \frac{\partial}{\partial k^{\rm un}_l} \left[{\bf Q}^{-1}_\mathcal{K}\right]_{ij} = \left[{\bf Q}^{-1}_\mathcal{K}\right]_{il}\left[{\bf Q}^{-1}_\mathcal{K}\right]_{lj}.
\end{equation}

\end{document}